%% file: 00-main.tex
\documentclass{easychair}

\title{Experiments in Verification of Linear Model Predictive Control:
  Automatic Generation and Formal Verification of an Interior Point
  Method Algorithm\thanks{This work was partially supported by ANR FEANICSES project.}}

 \author{Guillaume Davy\inst{1}\inst{2}
     \and
     Eric Feron\inst{3}
   \and
     Pierre-Loic Garoche\inst{1}
     \and
     Didier Henrion\inst{2}}
 
   \institute{Onera - The French Aerospace Lab, Toulouse, FRANCE \and CNRS LAAS, Toulouse, FRANCE \and Georgia Institute of Technology, Atlanta GA, USA}

\authorrunning{Davy, Feron, Garoche, Henrion}

\titlerunning{Verification of an IPM for linear MPC}

\input{01-macros}

\begin{document}

\maketitle


\begin{abstract}
Classical control of cyber-physical systems used to rely on basic
linear controllers. These controllers provided a safe and robust
behavior but lack the ability to perform more complex controls such as
aggressive maneuvering or performing fuel-efficient controls. Another
approach called optimal control is capable of computing such difficult
trajectories but lacks the ability to adapt to dynamic changes in the
environment. In both cases, the control was designed offline, relying
on more or less complex algorithms to find the appropriate
parameters. More recent kinds of approaches such as Linear
Model-Predictive Control (MPC) rely on the online use of convex
optimization to compute the best control at each sample time. In these
settings optimization algorithms are specialized for the specific
control problem and embed on the device.

This paper proposes to revisit the code generation of an interior
point method (IPM) algorithm, an efficient family of convex
optimization, focusing on the proof of its implementation at code
level.  Our approach relies on the code specialization phase to
produce additional annotations formalizing the intented specification
of the algorithm. Deductive methods are then used to prove
automatically the validity of these assertions. Since the algorithm is
complex, additional lemmas are also produced, allowing the complete
proof to be checked by SMT solvers only.

This work is the first to address the effective formal proof of an IPM
algorithm. The approach could also be generalized more systematically to code
generation frameworks, producing proof certificate along the code, for
numerical intensive software.





\end{abstract}


\input{1-introduction}
\input{2-prelim}
\input{3-code}
\input{5-proof}
\input{6-results}
\input{7-related}
\vspace{-1em}
\input{8-conclusion}



\bibliographystyle{splncs}
\bibliography{main}

\end{document}

%% file: 01-macros.tex
\usepackage{etex}
\usepackage[english]{babel}
\usepackage[T1]{fontenc}
\usepackage[utf8]{inputenc}
\usepackage{amsmath,amsfonts,amssymb}
\usepackage{stmaryrd} 
\usepackage{xspace}
\usepackage{listings}

\usepackage{newlfont}
\usepackage{multicol}
\usepackage{graphicx}
\usepackage{comment}
\usepackage{url}
\usepackage[noend]{algorithmic}
\usepackage{algorithm}
\usepackage{hyperref}
\usepackage{cancel}
\usepackage{paralist}
\usepackage{bussproofs}
\usepackage{wrapfig}
\usepackage{bm}
\usepackage{booktabs}
\usepackage{tabularx}

\usepackage{pgfplots,tikz}
\definecolor{ffwwqq}{rgb}{1.,0.4,0.}
\definecolor{qqttzz}{rgb}{0.,0.2,0.6}
\definecolor{qqzzff}{rgb}{0.,0.6,1.}
\definecolor{ffccww}{rgb}{1.,0.8,0.4}
\definecolor{ruruzw}{rgb}{0.0784313725490196,0.0784313725490196,0.5882352941176471}
\definecolor{qqqqff}{rgb}{0.,0.,1.}
\definecolor{ffqqqq}{rgb}{1.,0.,0.}
\definecolor{ttzzqq}{rgb}{0.2,0.6,0.}
\definecolor{qqwwtt}{rgb}{0.,0.4,0.2}

\newtheorem{mydef}{Definition}

\newtheorem{myth}{Theorem}
\newtheorem{remark}{Remark}

\newcommand{\hoare}[3]{\textnormal{$\{#1\}$~\texttt{#2}~$\{#3\}$}}

\newcommand{\code}[1]{\textnormal{\texttt{#1}}}

\newcommand{\nlocal}[2]{\| #1 \|^*_{#2}}

 \lstdefinestyle{cacsl}{language=C++,
  alsoletter={<==>,==>,\&\&,||,\\,->},
  keywords = [2]{assumes,requires,ensures,assigns,behavior,logic,axiomatic,type,predicate,axiom,<==>,==>,\&\&,||,assert,loop-invariant, variant, ghost, for},
  keywords = [3]{\\valid,\\separated,\\forall,\\result,\\exact,\\model,\\total_error,\\error,\\abs,\\cos,\\sin,\\set_model,\\sqrt, \\old, reads},
  keywords = [4]{mat_gt, mat_mult, MatVar, getM, getN, mat_get, mat_scal, inv, mat_add, transpose, MatCst_1_1, MatCst_2_3},
  keywords = [5]{integer, LMat, real},
  keywordstyle=[1]\color{blue},
  keywordstyle=[2]\color{Purple},
  keywordstyle=[3]\color{BlueGreen},
  keywordstyle=[4]\color{BurntOrange},
  keywordstyle=[5]\color{ForestGreen},
  morecomment=[l][\color{lightblue}]{//},
  morecomment=[s][\color{lightblue}]{/*}{*/},
  moredelim=**[s][\color{OliveGreen}]{/*@}{*/},
  moredelim=**[l][\color{OliveGreen}]{//@},
  rulecolor=\color{black},
  alsoother = ,}
  
\lstdefinestyle{acsl}{language=C++,
  alsoletter={<==>,==>,\&\&,||,\\,->},
  basicstyle=\color{OliveGreen}\linespread{1.0}\scriptsize\ttfamily,
  keywords = [2]{assumes,requires,ensures,assigns,behavior,logic,axiomatic,type,predicate,axiom,<==>,==>,\&\&,||,assert,loop-invariant, variant, ghost, for},
  keywords = [3]{\\valid,\\separated,\\forall,\\result,\\exact,\\model,\\total_error,\\error,\\abs,\\cos,\\sin,\\set_model,\\sqrt, \\old},
  keywords = [4]{mat_gt, mat_mult, MatVar, getM, getN, mat_get, mat_scal, inv, mat_add, transpose, MatCst_1_1, MatCst_2_3},
  keywords = [5]{integer, LMat, real},
  keywordstyle=[1]\color{blue},
  keywordstyle=[2]\color{Purple},
  keywordstyle=[3]\color{BlueGreen},
  keywordstyle=[4]\color{BurntOrange},
  keywordstyle=[5]\color{ForestGreen},
  rulecolor=\color{black},
  alsoother = ,}

\lstdefinelanguage{Why3}
{
basicstyle=\ttfamily,
keywordstyle=\color{blue},
morekeywords=[1]{inductive,predicate,function,goal,type,use,import,theory,end,in,axiom,lemma,export,forall,constant,module,let,exception,match,with,exists,val,unit},%
morekeywords=[2]{true,false},%
otherkeywords={},%
commentstyle=\itshape,%
columns=[l]fullflexible,%
sensitive=true,%
morecomment=[s]{(*}{*)},%
escapeinside={*?}{?*},%
keepspaces=true,
literate=%
{v1}{v$_1$}{1}%
{v2}{v$_2$}{1}%
{v3}{v$_3$}{1}%
{<}{$<$}{1}%
{>}{$>$}{1}%
{<=}{$\le$}{1}%
{>=}{$\ge$}{1}%
{<>}{$\ne$}{1}%
{/\\}{$\land$}{1}%
{\\/}{ $\lor$ }{3}%
{\ or(}{ $\lor$(}{3}%
{not\ }{$\lnot$ }{1}%
{not(}{$\lnot$(}{1}%
{->}{$\rightarrow$}{2}%
{<->}{$\leftrightarrow$}{2}%
}


%% file: 1-introduction.tex
\section{Model Predictive Control and Verification Challenges}
\label{sec:intro}

When one wants to control the behavior of a physical device, one could
rely on the use of a feedback controller, executed on a computer, to
perform the necessary adjustements to the device to maintain its state
or reach a given target. Classical means of this \emph{control
  theory} amount to express the device behavior as a linear ordinary
differencial equation (ODE) and define the feedback controller as a
linear system; eg. a PID controller. The design phase searches for
proper \emph{gains}, ie. parametrization, of the controller to achieve
the desired behavior.

While this approach has been used for years with great success, eg. in
aircraft control, some more challenging behaviors or complex devices
need more sophisticated controllers. Assuming that the device behavior
is known, one can predict its future states. A first approach,
\emph{Optimal Control} with indirect-method, search for optimal
solutions solving a complex mathematical problem, the Pontryagin
Maximal Principle. This is typical used to compute rocket or
satellite trajectories. However this approach, while theoretically
optimal, requires complex computation and cannot yet be performed
online, in real-time. A second approach, \emph{Linear Model Predictive
  Control} or direct method, amounts to solve online a convex
optimization problem describing the current state, the target and the
problem constraints, ie. limits on the thrust. A pregnant example of
such trajectory computation is the landing of SpaceX orbital
rockets~\cite{lars}. 

The use of a convex encoding of the problem guarantees the absence of
saddle points (local minimums) and could be resolved efficiently with
polynomial-time algorithms. Convex optimization covers a large set of
convex-conic problems, from linear programming (LP: linear
constraints, linear cost) to quadratic programming (QP: linear
contraints, quadratic cost) or semi-definite programming (SDP: linear
constraints over matrices, linear cost). While the famous Simplex
algorithm can efficiently address LP ones, the Interior Point Method
(IPM) is the state-of-the art one when it comes to more advanced cones
(eg. QP, SDP).

Linear Model Predictive control can then be expressed as a
bounded model-checking problem with an additional cost to find the
best solution, using convex optimization:

\begin{mydef}[Example: LP encoding of MPC]
  Let $U \subseteq \mathbb{R}^u$ and $X \subseteq \mathbb{R}^s$ be
  constrained convex set for inputs and states of the controller. Let
  $(u_k \in U)_{0 \leq k < N}$ be an $N$-bounded sequence of control
  inputs for a linear system $x_{k+1} = A x_k + B u_{k}$ with $x_k \in
  X$ a vector of state variables, $A \in \mathbb{R}^{s \times s}$ and
  $B \in \mathbb{R}^{s\times u}$. Let $X_0$ the initial state and
  $X_N$ the target one. The objective is to compute an optimal
  trajectory, for example minimizing the required inputs
  $\Sigma_{i=1}^N |u_i|$ to reach the target point $X_N$. Let us
  define the following LP problem:
\begin{equation}
  \label{eq:mpc}
  \min \Sigma_{i=1}^N |u_i|\\
\textrm{ s.t. } \left\{
\begin{array}{l}
  X_0 = x_0, X_N = x_n,\\
x_1 = A \cdot x_0 + B \cdot u_0,\\
\vdots\\
x_N = A \cdot x_{N-1} + B \cdot u_{N-1}.\\
\end{array}\right.
\end{equation}
\end{mydef}

Let us remark that Eq.~\eqref{eq:mpc} relies on a linear and discrete
description (through matrices $A$ and $B$) of the original device
behavior, typically a non linear ODE. This LP problem has $(N-2)
\times s + N\times u$ variables since $x_0, x_N, A$ and $B$ are
known.
Without loss of generality, one can express Eq.~\eqref{eq:mpc} over fresh definitions of $A \in \mathbb{R}^{m \times n}$, $b \in \mathbb{R}^m$ and $c \in \mathbb{R}^n$ as the following LP:
\begin{equation}
  \label{eq:lp}
  \begin{array}{l}
    \min \; c^\intercal \cdot x\\
    \textrm{ s.t. } A \cdot x \leq b
    \end{array}
\end{equation}
with $m$ constraints, and where $x \in \mathbb{R}^n$ denotes here a
larger state. Equality constraints $x = e$ are defined as $x \leq e
\land -x \leq -e$.

This computation is embedded on the device, and, depending on the
starting point $X_0$, computes the sequence of $N$ inputs to reach the
final point $X_N$. Since most of the parameters are known \emph{a
  priori} the \emph{usual approach is to instanciate an optimization
  algorithm to the specific problem} rather than embed on the device a
generic solver. For example SpaceX rockets rely on
CVX~\cite{cvxboyd2014} to produce custom code, compatible with
embedded devices (eg. no dynamic allocation, reduced number of
computations).

While classical linear control only amounts to the computation of a
linear update, these MPC approaches rely on more involved online
computation: the convex optimization has to return a sound solution
in finite time.

The objective of this paper is to address the verification of these
IPM implementations to ensure the good behavior of MPC-based
ccontrollers.
A possible approach could have been to specify the algorithm in a
proof assistant and extract a correct-by-construction
code. Unfortunately code extraction from these proof assistants
generates source code that hardly resembles classical critical
embedded source code. We rather chose to synthesize the final code
while producing automatically code contracts and additional
annotations and lemmas to support the automatic validation of the
code.
Our contributions are then following:
\begin{enumerate}
\item We revisit the custom code generation, instanciating an IPM
  algorithm to the provided problem, producing both the code, its
  formal specification~\cite{hoare69,Floyd1967Flowcharts} and proof
  artifacts.
\item We then rely on Frama-C~\cite{frama-c} to prove the validity of
  generated Hoare triples using deductive methods~\cite{dijkstra75}.
\item Our proof process is automatic: instead of proving a generic
  version of the algorithm, which would require strong assumptions on
  the problem parameters, we perform an automatic instance-specific
  proof, achieving a complete validation without any user input: all
  proofs are performed, at code level, using SMT-solvers only.
\item The approach has been evaluated on a set of generated instances
  of various sizes, evaluating the scalability of the proof
  process. The generated code can then be embedded in actual devices.
\end{enumerate}

The considered setting is a primal IPM solving a LP problem. In the
presented work, we focus only on the algorithmic part while dealing
with the actual implementation, therefore numerical issues such as
floating point computation are here neglected.

This work is the first approach addressing the formal verification of
an IPM convex algorithm, as used in MPC-based controllers.

This paper is structured as follow. We introduce the reader to convex
optimization and IPM in Sec.~\ref{sec:prelim} and we describe the code
and specification generation in
Sec.~\ref{sec:code}. Sec.~\ref{sec:proof} focuses on the proof process
supporting the automatic proof of generated code specification.
Sec.~\ref{sec:results} provides a feedback on our approach
evaluation. Related works are then presented in
Sec.~\ref{sec:related}.



%% file: 2-prelim.tex
\section{Convex Optimization with the Interior Point Method}
\label{sec:prelim}

Let us outline the key definition and principles behind the primal IPM
we analyze. As mentioned above, these notions easily extend to more
sophisticated cones. We chose a primal algorithm for its simplicity compared to
primal/dual algorithms and we followed Nesterov's proof~\cite{nesterov94}.

\begin{mydef}[LP solution]\label{def:lp_sol}
Let us consider the problem of Eq.~\eqref{eq:lp} and assume that an optimal point $x^*$ exists and is unique. We have then the following definitions:
\begin{align}
 E_{f} =& \{ x \in \mathbb{R}^n \mid A x < b \} & & \text{(feasible set of }P \text{)}\label{eq:feasible_set}\\
 f(x) =& \langle c, x \rangle = C^\intercal \cdot x && \text{(cost function)}\label{eq:cost_fcn}\\
 x^* =& \arg\min \limits_{x \in E_f} f&& \text{(optimal point)}\label{eq:opt_pt}
\end{align}
\end{mydef}

\begin{wrapfigure}{R}{0.405\textwidth}
  \vspace{-2em}
\centering
\includegraphics[width=0.4\textwidth]{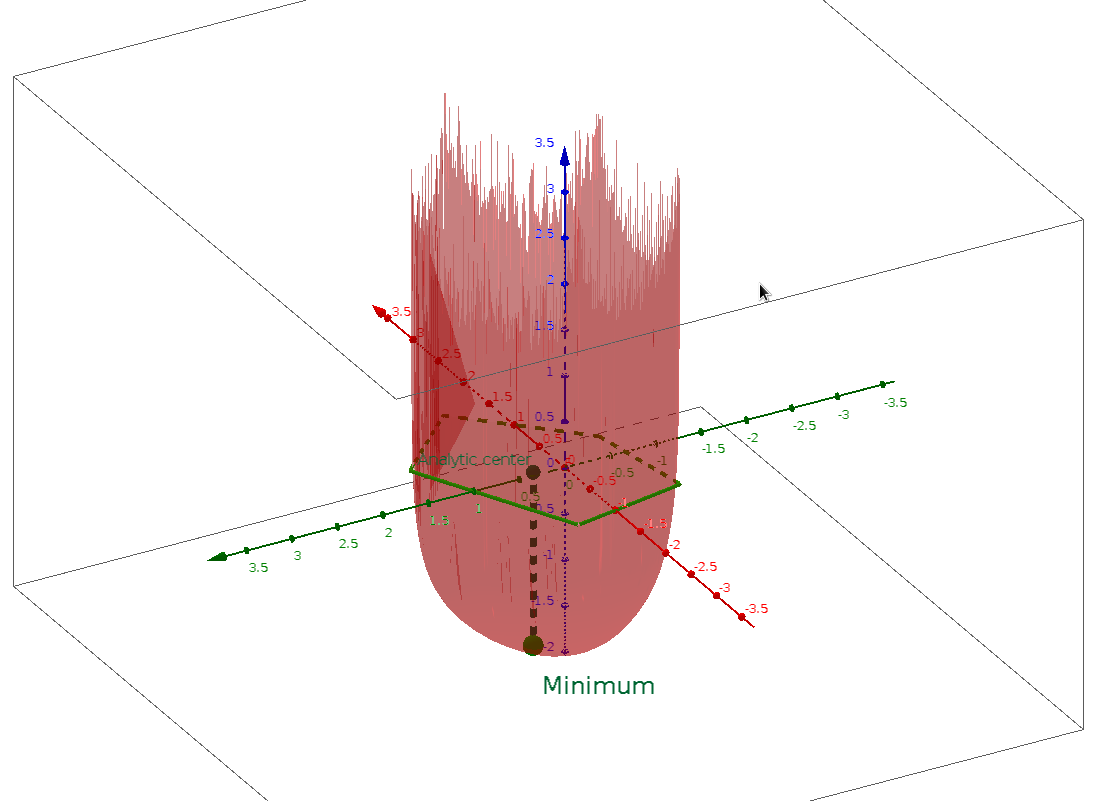}
\vspace{-1em}
\caption{Barrier function.}
\label{fig:barrier}
  \vspace{-4em}
\end{wrapfigure}
In order to ensure the existence of an optimal value, we assume the set $E_f$ to
be bounded.

\paragraph{Barrier function and analytic center.}
One can
describe the interior of the feasible set using a penalty function
$F:E_f \rightarrow \mathbb{R}$. Figure~\ref{fig:barrier} depicts such
a logarithmic barrier function encoding a set of linear constraints,
it diverges when $x$ approaches the border of $E_f$. By construction, it
admits a unique minimum, called the \emph{analytic center}, in the interior of the feasible set.

\paragraph{Central path}
A minimum of a function is obtain by analyzing the zeros of its
gradients; for convex function this minimum is unique. In case of a
cost function with constraints, IPM represents these constraints within
the cost function using the function $\tilde{f}(x,t)$. The variable
$t$ balances the impact of the barrier: when $t=0$, $\tilde{f}(x, t)$
is independent from the objective while when $t \rightarrow + \infty$,
the cost gains more weight.

\noindent\begin{minipage}{.65\textwidth}
\begin{mydef}[Adjusted cost $\tilde{f}$.] 
Let $\tilde{f}$ be a linear combination of the previous objective
function $f$ and the barrier function $F$.
\begin{equation}
\tilde{f}(x, t) = t \times f(x) + F(x) \text{ with } t \in \mathbb{R}
\end{equation}
\label{eq:new_cost}
\end{mydef}

\begin{mydef}[Central path: from analytic center to optimal solution.]
The values of $x$ minimizing $\tilde{f}$
when $t$ varies from $0$ to $+ \infty$ characterize a
path, the \textit{central path} (cf. Fig.~\ref{fig:central_path}). 
\begin{equation}
\begin{aligned}
x^* : ~~ & \mathbb{R}^+&\rightarrow& ~ E_{f}\\
         & t           &\mapsto    & ~ \arg\min \limits_{x \in E_{f}} \tilde{f}(x, t)
\end{aligned}
\end{equation}
Note that $x^*(0)$ is the analytic center while $\lim \limits_{t \rightarrow + \infty} x^*(t) = x^*$.
\end{mydef}
\end{minipage}%
\begin{minipage}{.35\textwidth}
\begin{figure}[H]
  \centering
  \vspace{3em}
\includegraphics[width=.65\textwidth]{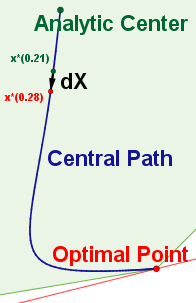}
\caption{Central path.}
\label{fig:central_path}
\vspace{0em}
\end{figure}
\end{minipage}

\paragraph{IPM algorithm: a sequence of Newton steps following the central path.}
IPM algorithm performs a sequence of iterations, updating a point $X$ that
follows the central path and eventually reaches the optimal point. At the beginning
of each loop iteration, there exists a real $t$ such that $X = x^*(t)$. Then $t$ is increased
by $dt > 0$ and $x^*(t+dt)$ is the new point $X$. This translation
$dX$ is performed by a Newton step. 

\begin{paragraph}{Computing $dX$ using Newton step.}
We recall that 
the Newton's method computes an approximation of a root of a function
$G:\mathbb{R}^k \rightarrow \mathbb{R}^l$.  It is a first order method, ie. it
relies on the gradient of the function and, from a point in the domain of the
neighborhood of a root, performs a sequence of iterations, called Newton
steps. 
A Newton step transforms a point $Y_n$ into $Y_{n+1}$ as follows:
\begin{equation}
Y_{n+1} - Y_{n} = -{\bigl (}G'(Y_{n}){\bigr )}^{-1}G(Y_{n}) 
\end{equation}
\end{paragraph}


We apply the Newton step to the gradient of $\tilde{f}$, computing its root
which coincides with the minimum of $\tilde{f}$. We obtain:
\begin{equation}\label{eq:dX}
dX = -{\bigl (}F''(X){\bigr )}^{-1}((t + dt)c + F'(X)) 
\end{equation}

\paragraph{Computing dt: preserving the Approximate Centering Condition (ACC).}
The convergence of the Newton method is
guaranteed only in the neighborhood to the function root. This
neighborhood is called the region of quadratic convergence; this
region evolves on each iteration since $t$ varies.
To guarantee that
the iterate $X$ remains in the region after each iteration, we require
the barrier function to be self-concordant.  Without providing the
complete definition of self-condordance, let us focus on the implied
properties:
assuming that $F$ is a self-concordant, then  
%
%
%
$F''$, its Hessian, is non-degenerate(\cite[Th4.1.3]{nesterov94}) and
we can define a local norm.

\begin{mydef}[Local-norm]
\label{def:norm}
\begin{equation}\label{eq:norm}
\nlocal{y}{x} = \sqrt{y^T \times F''(x)^{-1} \times y}
\end{equation}

\end{mydef}

This local-norm allows one to define the Approximate Centering Condition (ACC),
the crucial property which guarantees that $X$ remains in the region
of quadratic convergence:

\begin{mydef}[ACC]
\label{def:acc}
Let $x \in E_f$ and $t \in \mathbb{R}^+$, $ACC(x, t, \beta)$ is a predicate defined by
\begin{equation}
\nlocal{\tilde{f}'(x)}{x} = \nlocal{tc + F'(x)}{x} \leq \beta
\end{equation}
\end{mydef}

In the following, as advised in~\cite{nesterov94}, we choose a specific value for $\beta$, as defined in~\eqref{eq:beta_cond}. 

\begin{equation}\label{eq:beta_cond}
\beta < \frac{3 - \sqrt{5}}{2}
\end{equation}

The only step remaining is the computation of the largest $dt$
such that $X$ remains in the region of quadratic convergence around $x^*(t + dt)$, with $\gamma$ a constant:
\begin{equation}\label{eq:dt}
dt = \frac{\gamma}{\nlocal{c}{x}}
\end{equation}

\begin{myth}[ACC preserved]
\label{th:acc}
This choice maintains the ACC at each iteration(\cite[Th4.2.8]{nesterov94}). When $ACC(X, t, \beta)$ and
$\gamma \leq \frac{\sqrt{\beta}}{1 + \sqrt{\beta}} - \beta$
then $ACC(X+dX, t+dt, \beta)$.
\end{myth}

\paragraph{Summary of IPM}
Thanks to an appropriate barrier function to describe the
feasible set, IPM algorithm starts from the point $X = x^*(0)$, the analytic
center, and $t = 0$. Then its updates both variables using
Eqs.~\eqref{eq:dX} and~\eqref{eq:dt}, until the required precision is reached.

\begin{remark}[Choice of a barrier function.]
For this work, we use the classic self-concordant barrier for linear
programs:
\begin{equation}
F(x) = \sum \limits_{i=0}^{m} - log(b_i - A_i \times x)\label{eq:bf}
\end{equation}
with $A_1 , A_m$ the columns of $A$.
\end{remark}

\begin{remark}[Computation of the analytic center.]
\label{par:AC}
$x^*_F$ is required to initiate the algorithm. In case of offline use
the value could be precomputed and validated. However in case of
online use, its computation itself has to be proved. Fortunately this
can be done by a similar algorithm with comparable proofs. In
addition, in MPC-based controllers, the set of constraints can be
fed with previously computed values, guaranteeing the existence of a non-empty
interior and provding a feasible point to compute this new analytic
center.
\end{remark}

%% file: 3-code.tex
\section{Generating Code and Formal Specification}
\label{sec:code}

Typical uses of MPC do not rely on a generic solver implementing an
IPM algorithm with a large set of parameters but rather instanciate
the algorithm to the provided instance. As mentioned in the
introduction, a large subset of the problem is known beforehand and
therefore lots of computation can be hard-coded. As an example, the
computation of the local norm relies mainly on the computation of the
Hessian of the barrier function and can be, in some cases,
precomputed.

The code generation is therefore very similar from one instance to the
other. The main changes lie in the sizes of the various arrays
representing the problem and the points along the central path.

\begin{wrapfigure}{R}{0.505\textwidth}
  \centering
\includegraphics[width=0.4\textwidth]{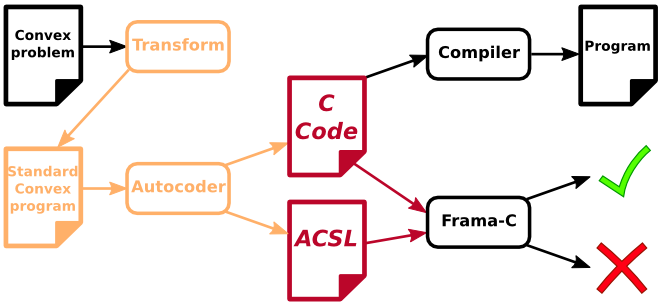}
\caption{Autocoding toolchain: automatic generation of both C code and annotations.}
\label{fig:toolchain}
\end{wrapfigure}
Figure~\ref{fig:toolchain} sketches our fully automatic process which,
when provided with a convex problem with some unknown values,
generates the code, the associated annotation and prove it
automatically. We are not going to present all the process in this
paper but concentrate on how to write the embedded code, annotate it
and automatize its proof. Early stages of the process reformulate the
given instance in a canoncial form. Regarding the code generation, our
approach is really similar to the CVXGEN or CVXOPT tools.

\subsection{Code structure: Newton steps in a For-Loop}
The generated code follows strictly the algorithm presented in the
previous section for each iteration steps. However, usual
implementations perform the iterations in a while-loop until a
condition $t_k > t_{stop}(\epsilon)$ is satisfied, where $t_k$
represents the position on the central path at iteration $k$.

In order to guarantee termination, we rely on the convergence proof
and complexity analysis of~\cite[Th 4.2.9]{nesterov94} and compute, a
priori, the required number of iterations $k_{last}(\epsilon)$ to
reach a given optimality $\epsilon$.  The termination proof relies on
the characterization of a a geometric progression $\textsc{Lower}(k)$
such that each increment $dt$ is sufficient to achieve some progress
towards termination: at $k$-th iteration, we have
\begin{equation}\label{eq:lower}
t_k \geq \textsc{Lower}(k)
\end{equation}

In terms of practical impact the code becomes a for loop with a
provided number of iterations while the $\textsc{Lower}$ progression
only appears in the formal specification part and is used for proof
purposes.

\subsection{Domain-specific axiomatics}

We rely on the tool Frama-C~\cite{frama-c} to perform the proofs at
code level and on ACSL~\cite{acsl}, its annotation language, to
formally express the specification. While extensible, ACSL does not
provide high level constructs regarding linear algebra or optimzation
related properties. We first defined new ACSL axiomatics: specific
sets of types, operators and axioms to express these. Similar
approaches were already proposed~\cite{nfm12_2} for matrices but were
too specific to ellipsoid problems.  We give here an overview of the
definitions of both the Matrix and Optim ACSL axiomatics, both presented in Fig.~\ref{fig:axiomatics}. Note that
these definition are not instance-specific and are shared among all
uses of our framework.

Axiomatic are defined as algebraic specifications. Logical opererators manipulating the local types can either be defined as functions or as uninterpreted functions fitted with axioms.

\paragraph{Matrix axiomatic.}
The new ACSL type \code{LMat}, ie. Logic Matrix, is
introduced and fitted with getters and constructors. Constructors such
as \code{MatVar} allow to bind fresh \code{LMat} value. This one reads
a C array while others create constant values. Operators such as matrix addition or inverse are defined axiomatically: the operator is defined and fitted with axioms. 

\begin{figure}
\begin{minipage}{.5\textwidth}
\begin{lstlisting}[style=acsl]
axiomatic matrix
{
	type LMat;
        // Getters
	logic integer getM(LMat A);
	logic integer getN(LMat A);
	logic real mat_get(LMat A, integer i, integer j);
        // Constructors
	logic LMat MatVar(double* ar, integer m, integer n) reads ar[0..(m*n)];
	logic LMat MatCst_1_1(real x0);
	logic LMat MatCst_2_3(real x0, real x1, real x2, real x3, real x4, real x5);
        // Operators
	logic LMat mat_add(LMat A, LMat B);
	logic LMat mat_mult(LMat A, LMat B);
	logic LMat transpose(LMat A);
	logic LMat inv(LMat A);
	...
        // Example of axioms
	axiom getM_add:	\forall LMat A, B; getM(mat_add(A, B))==getM(A);
	axiom mat_eq_def:
	\forall LMat A, B;
		(getM(A)==getM(B))==> (getN(A)==getN(B))==>
		(\forall integer i, j; 0<=i<getM(A) ==> 0<=j<getN(A) ==>
			mat_get(A,i,j)==(mat_get(B,i,j))==>
			A == B;
	...
}
\end{lstlisting}
\end{minipage}%
\begin{minipage}{.5\textwidth}
\begin{lstlisting}[style=acsl]
axiomatic Optim
{
	logic LMat hess(LMat A, LMat b, LMat X);
	logic LMat grad(LMat A, LMat b, LMat X);
	logic real sol(LMat A, LMat b, LMat c);
        logic real sol(LMat A, LMat b, LMat c);
          axiom sol_min: \forall LMat A, b, c;
	    \forall LMat y; mat_gt(mat_mult(A, y), b) ==>
		dot(c, y) >= sol(A, b, c);
          axiom sol_greater: \forall LMat A, b, c;
	    \forall Real y;
		(\forall LMat x; mat_gt(mat_mult(A, x), b) ==> dot(c, x) >= y) ==>
			sol(A, b, c) >= y;
	logic real norm(LMat A, LMat b, LMat x, LMat X) =
		\sqrt(mat_get(mat_mult(transpose(x), mat_mult(inv(hess(A, b, X)), x)), (0), (0)));
	logic boolean acc(LMat A, LMat b, LMat c, real t, LMat X, real beta) =
		((norm(A, b, mat_add(grad(A, b, X), mat_scal(c, t)), X))<=(beta));
	...
}
\end{lstlisting}
\end{minipage}
\vspace{-3em}
\caption{ACSL Axiomatics: Matrix and Optim}
\label{fig:axiomatics}
\end{figure}

\paragraph{Optimization axiomatic.}
Besides generic matrix operators, we also need some operators specific
to our algorithm. The Newton step defined in Eq.~\eqref{eq:dX} as
well as the local norm of Eq.~\eqref{eq:norm} are both based on the
gradient and the Hessian of the barrier function (Eq.~\eqref{eq:bf})
encoding the feasible set. These functions are parametrized by the
matrix $A$, the vector $b$ and the local point $X$.  However Hessian
and gradient are hard to define without real analysis which is well
beyond the scope of this article. Therefore we decided to directly
axiomatize some theorems relying on their definition like
\cite[Th4.1.14]{nesterov94}.

Another important notion is the optimal solution $x^*$ of the convex
problem. We can characterize axiomatically the
equations~\eqref{eq:feasible_set} -- \eqref{eq:opt_pt} of
Definition~\ref{def:lp_sol} using the uninterpreted function
\code{sol}:
\begin{equation}
\begin{array}{lcl}
\forall y \in E_f, c^T y \geq \code{sol} &&
\forall y \in \mathbb{R}, \forall x \in E_f, c^T x \geq y \implies \code{sol} \geq y
\end{array}\end{equation}

Last we defined some operators representing definitions
\ref{def:norm}, \ref{def:acc} and \textsc{Lower}\eqref{eq:lower}.

\subsection{Functions and contracts}

\paragraph{Multiple functions to support proof and local contrats.}
Proving large functions is usually hard with SMT-based reasoning since the
generated goals are too complex to be discharged automatically. A more efficient
approach is to associate small pieces of code with local contracts; these
intermediate annotations acting as cut-rules in the proof processes.

Let \code{A = B[C]} be a piece of code containing \code{C}. Replacing \code{C}
by a call to \code{f() \{ C \}} requires either to inline the call or to write a
new contract \hoare{P}{f()}{Q}, characterizing two smaller goals instead of a
larger one.  Specifically in the proof of a \code{B[f()]}, \code{C} has been
replaced by $P$ and $Q$ which is simpler than an automatically computed weakest
precondition.

\begin{figure}
\vspace{-1em}
\begin{lstlisting}[style=cacsl]
/*@ ensures MatVar(dX, 2, 1) == \old(mat_scal(MatVar(cholesky, 2, 1), -1.0));
  @ assigns *(dX+(0..2)); */
void set_dX()
{
    dX[0] = -cholesky[0];      dX[1] = -cholesky[1];
}
\end{lstlisting}
Simple case of \code{dx \%= -cholesky} encapsulated in a function for
\code{dx} and \code{cholesky} of size $2 \times 1$.

The first \textit{ensures} statement specifies, using our
own encoding of matrices in ACSL, that, after the execution of the function
body, the relationship $dX = -1 * cholesky$ holds where both $dX$ and $cholesky$
are interpreted as matrices.  The second annotation \textit{assigns} is used for
the alias analysis and expresses that the only modified values are
$dX[0]$, $dX[1]$. This property is also automatically generated and has to be
proved by the solver. It is of great help to support the proof of relationships
between arrays.
Note
that all variables are global here.
\caption{Example of ACSL-specified matrix operation.}
\label{fig:mat_op}
\end{figure}

Therefore instead of having one large function, our code is structured
into several functions. As an example, each basic matrix operation is
performed in a dedicated function. The associated contract provides a
high level interpretation of the function behavior expressed over
matrices abstract datatypes. This modular encoding supports the
generation of simpler goals for the SMT solver, while keeping the code
structure clearer.

\begin{figure}
\includegraphics[width=\textwidth]{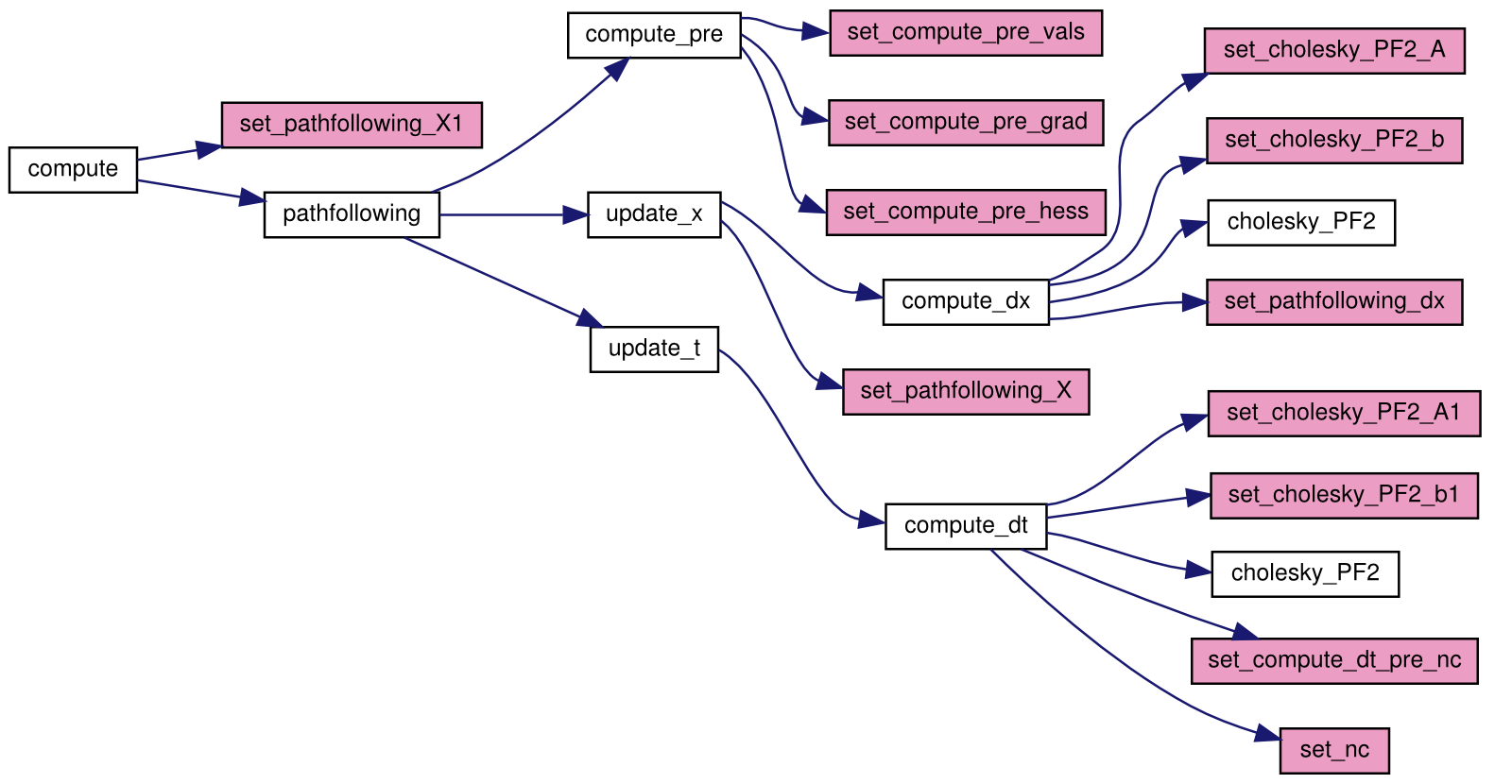}

\vbox{\vspace{-6cm}\scriptsize\begin{itemize}
	\item \code{compute} fill $X$ with the the analytic\\
	center and call \code{pathfollowing}.
	\item \code{pathfollowing} contains the main loop which\\
	udpate $dX$ and $dt$.
	\item \code{compute\_pre} computes Hessian and gradiant of\\
	$F$ which are required for $dt$ and $dX$.
	\item \code{udpate\_dX} and \code{udpate\_dt} call the
          associated subfunction and\\ update the corresponding value.
	\item \code{compute\_dt} performs~\eqref{eq:dt}, it requires
          to call Cholesky to compute\\ the local norm of $c$.
	\item \code{compute\_dX} performs~\eqref{eq:dX}, Cholesky is
          used to inverse the Hessian matrix.
\end{itemize}}
\caption{Call tree of the implementation (colored boxes are matrix computation)}
\label{fig:calltree}
\end{figure}

Fig.~\ref{fig:mat_op} illustrates a very simple generated code and
contract while Fig.~\ref{fig:calltree} presents the hierarchy of function calls.

\paragraph{Specification of \code{pathfollowing}.}

A sound algorithm must produce a point in the feasible set such that
its cost is $\epsilon$-close to $sol$. This represents the functional
specification
that we expect from the code and is asserted by two post-conditions.
In addition, two preconditions state that $X$ is feasible and close enough to the analytic center. 


\begin{wrapfigure}{r}{.5\textwidth}
 \vspace{-1.6em}
\begin{lstlisting}[style=cacsl]
/*@ requires mat_gt(mat_mult(A, MatVar(X, N, 1)), b);
  @ requires acc(A, b, c, 0, MatVar(X, N, 1), BETA);
  @ ensures mat_gt(mat_mult(A, MatVar(X, N, 1)), b);
  @ ensures dot(MatVar(X, 2, 1), c) - sol(A, b, c) < EPSILON
*/
void pathfollowing() {
  ...
  /*@ loop-invariant mat_gt(mat_mult(A, MatVar(X, N, 1)), b);
    @ loop-invariant acc(A, b, c, t, MatVar(X, N, 1), BETA);
    @ loop-invariant t > lower(l); */
  for (int l = 0; l < NBR;l++) { ... }
  ...
}
\end{lstlisting}
\vspace{-1em}
\caption{\code{Pathfollowing} ACSL contract and loop annotations.}
\label{fig:acsl_pathfollowing}
\vspace{-1em}
\end{wrapfigure}Thanks to our two new theories \code{Matrix} and \code{Optim}, writing
and reading this contract is straightforward and can be checked by
anyone familiar with linear programming.
Our contribution
focuses on the expression on intermediate
annotations, spread over the code, to support
the overall proof of
specifications.

A loop needs to be annotated by an invariant to have its Weakest precondition
computed. We need three invariants for our pathfollowing algorithm.
The first one guarantees the feasibility of $X$ while the second one
states the conservation of the $ACC$ (cf. Def.~\ref{def:acc})
The third invariant assert that $t$ is increasing enough on each iteration, more
specifically that it is greater than a geometric progression\eqref{eq:lower}.

Proving the initialization is straightforward, thanks to the main
preconditions.
We wrote one ACSL lemma for each invariant. Whenever it is possible we
try to follow \cite{nesterov94}, for example by translating
\cite[Th4.1.5]{nesterov94} and Theorem \ref{th:acc}.
The last two loop invariants are combined to prove the second
post-condition of \code{pathfollowing} thanks to Theorem
\ref{th:main}. \code{NBR} is computed from $k_{last}$.

\begin{myth}\label{th:main}\cite[Th4.2.7]{nesterov94}
Let $t \geq 0$, and $X$ such that $ACC(X, t, \beta)$ then
\begin{equation}
c^T X - c^T X^* < \frac{1}{t} \times (1 + \frac{(\beta + 1)\beta}{1 - \beta})
\end{equation}
\end{myth}

Fig.~\ref{fig:acsl_pathfollowing} presents the specification in ACSL:
the main function contract and the loop invariants.

\begin{wrapfigure}{r}{.5\textwidth}
  \vspace{-2em}
\begin{lstlisting}[style=cacsl]
/*@ requires MatVar(hess, N, N)==hess(A, b, MatVar(X, N, 1));
  @ requires acc(A, b, c, t, MatVar(X, N, 1), BETA);
  @ ensures  acc(A, b, c, t, MatVar(X, N, 1), BETA + GAMMA);
  @ ensures t > \old(t)*(1 + GAMMA/(1 + BETA));*/
void update_t();
\end{lstlisting}
\caption{\code{update\_t} ACSL contract}
\label{fig:update_t}
\end{wrapfigure}
\paragraph{Loop body.}

Each iteration of the IPM relies on three function calls:
\code{update\_pre} computing some common values, \code{update\_t} and
\code{update\_x} (cf Fig.~\ref{fig:calltree}). Theorem~\ref{th:acc} is then split into several properties and the
corresponding post-conditions. For example, the contract of
\code{update\_t} is described in Fig.~\ref{fig:update_t}.

The first post-condition is an intermediary results stating that:

\begin{equation}\label{eq:update_t_ensures_1}
ACC(X, t+dt, \beta + \gamma)
\end{equation}

This result is used as precondition for \code{update\_x}.  The second
\code{ensures} corresponds to the product of $t$ by the common ratio
of the geometric progression $\textsc{Lower},$ cf. Eq.~\eqref{eq:lower}
which will be used to prove the second invariant of the loop.  The
first precondition is a post-condition from \code{update\_pre} and the
second one is the first loop invariant.


%% file: 5-proof.tex
\section{Automatic Verification: Proof Refinement through Lemma Definitions and Local Contracts.}
\label{sec:proof}

Our framework produces both the code and its annotations, as well as a
set of axioms related to our newly defined axiomatics
(cf. Fig.~\ref{fig:axiomatics}). However most of the contracts cannot
be proved automatically by an SMT solver. After a first phase in which
we tried to perform these proofs with Coq, we searched for more
generic means to automatize the proof.

\subsection{Refining the proofs.}
We performed the following steps in order to identify the appropriate set of additional assertions, local contracts or lemma to be introduce:
\begin{itemize}
\item we took a fully defined custom code for a linear problem and study means
  to prove its validity: a feasible, optimal solution, while guarantying the
  convergence of the algorithm.
\item these proofs were then manually generalized through the introduction of
  numerous intermediate lemmas as annotations in the code. These additional
  annotations enable the automatic proof of the algorithm using an off-the-shelf
  SMT solver Alt-Ergo~\cite{altergo}.
\item the custom code generation was extended to produce both the actual code
  and these annotations and function contracts, enabling both the generation of
  the embedded code and its proof of functional soundness.
\end{itemize}

Another approach is to refine the code into smaller components, as presented in Fig.~\ref{fig:calltree}. The encoding hides low-level operations to the rest of the code, leading to two
kinds of goals:
\begin{itemize}
\item Low level operation (memory and basic matrix operation).
\item High level operation (mathematics on matrices).
\end{itemize}

\subsection{Simplifying the code: addressing memory-related issues.}


One of the difficulties when analyzing C code are memory related issues. Two
different pointers can \textit{alias} and reference the same part of the
stack. A fine and complex modeling of the memory in the predicate encoding, such
as separation logic\cite{1029817} could address these issues. Another more
pragmatic approach amounts to substitute all local variables and function
arguments with global static variables. Two static arrays cannot overlap since
their memory is allocated at compile time.

Since we are targeting embedded system, static variables will also permit to
compute and reduce memory footprints. However there are two major drawbacks: the
code is less readable and variables are accessible from any function. These
two points usually lead the programmer to mistakes but could be accepted in case
of code generation. In order to support this we also tagged all variables with
the function they belong to by prefixing each variable with its function name.

\subsection{Low-level goals: Proving Matrix operations.}

As explained in previous Section, the matrix computation are encapsulated into
smaller functions. Their contract states the equality between the resulting
matrix and the operation computed. The extensionality axiom (\code{mat\_eq\_def})
is required to prove this kind of contract. Extensionality means that if two
objects have the same external properties then they are equal.

This axiom belong to the matrix axiomatic but is too general to be used therefore
lemmas specific to the matrices size are added for each matrix affectation.
This lemma can be proven with the previous axioms and therefore does not introduce
more assumption.

The proof remains difficult or hardly automatic for SMT solvers therefore we
append additional assertions, as sketched in Figure~\ref{fig:mat_op_assert}, at
the end of function stating all the hypothesis of the extensionality lemma.
Proving these post-conditions is straightforward and smaller goals need now to be proven.

\begin{figure}
\begin{lstlisting}[style=acsl]
assert getM(MatVar(dX,2,1)) == 2;
assert getN(MatVar(dX,2,1)) == 1;
assert getM(MatVar(cholesky,2,1)) == 2;
assert getN(MatVar(cholesky,2,1)) == 1;
assert mat_get(MatVar(dX,2,1),0,0) == mat_get(\old(mat_scal(MatVar(cholesky,2,1),-1.0)),0,0);
assert mat_get(MatVar(dX, 2, 1),1,0) == mat_get(\old(mat_scal(MatVar(cholesky,2,1),-1.0)),1,0);
\end{lstlisting}
\caption{Assertion appended to the function of Figure \ref{fig:mat_op}}
\label{fig:mat_op_assert}
\end{figure}

Further split of functions into smaller functions is performed when
functions become larger. 
As an example, for operations on
matrices of size $m^2$, there is $m^2$ lines of C code manipulating
arrays and logic. To avoid large goals involving array accesses for
SMT solvers when $m$ becomes greater than $10$, each operation is
encapsulated as a separate function annotated by the operation on a
single element:
\begin{lstlisting}[style=acsl]
mat_get(MatVar(dX,2,1),0,0) == mat_get(\old(mat_scal(MatVar(cholesky,2,1),-1.0)),0,0);
\end{lstlisting}

These generated goals are small enough for SMT. The post-condition of the
matrix operation remains large but the array accesses are abstracted away by all the
function call and therefore becomes provable by SMT solvers.

\subsection{High-level goals: IPM specific properties.}

Proving sophisticated lemmas with non trivial proof is hardly feasible
for an SMT solver. To support them, we introduced many intermediate
lemmas. At each proof step (variable expansion, commutation,
factorization, lemma instantiation, ...) a new lemma is automatically
introduced. With this method, SMT solvers manage to handle each small
step which eventually lead to the proof of the initial lemma.


For example, the post-condition~\eqref{eq:update_t_ensures_1} of
\code{update\_t} is not provable by \textsf{Alt-Ergo}. In order to
prove it, we introduce the lemma~\eqref{eq:ensures1} where $P_1$ is $ACC(X,
t, \beta)$, the precondition, and $P_2$ is $dt =
\frac{\gamma}{\nlocal{c}{x}}$ the performed update.

\begin{equation}\tag{\code{update\_t\_ensures1}}\label{eq:ensures1}
\forall x, t, dt;~ P_1 \Rightarrow P_2 \Rightarrow ACC(X, t + dt, \beta + \gamma)
\end{equation}

Proving the post-condition with SMT solvers is straightforward given
the previous lemma but again proving the lemma itself is beyond their
capabilities. However with the additional lemma \eqref{eq:ensures1_l0}
the proof becomes obvious to the SMT solver. Indeed the lemma is just
the expansion of the ACC definition.

\begin{equation}\tag{\code{update\_t\_ensures1\_l0}}\label{eq:ensures1_l0}
\forall x, t, dt;~ P_1 \Rightarrow P_2 \Rightarrow \nlocal{F'(X) + c(t + dt)}{x} \leq \beta + \gamma
\end{equation}

Step by step, we introduce new lemmas to prove the previous ones: e.g. to prove the goal \eqref{eq:ensures1_l0} we introduce the following three lemmas:

\begin{equation}\tag{\code{update\_t\_ensures1\_l3}}\label{eq:ensures1_l3}
\forall x, t, dt;~ P_2 \Rightarrow \nlocal{c \times dt}{x} = \gamma
\end{equation}

\begin{equation}\tag{\code{update\_t\_ensures1\_l2}}\label{eq:ensures1_l2}
\forall x, t, dt;~ P_1 \Rightarrow \nlocal{F'(X) + c \times t}{x} \leq \beta
\end{equation}

\begin{equation}\tag{\code{update\_t\_ensures1\_l1}}\label{eq:ensures1_l1}
\forall x, t, dt;~ P_1 \Rightarrow P_2 \Rightarrow \nlocal{F'(X) + c(t + dt)}{x} \leq \nlocal{F'(X) + c \times t)}{x} + \nlocal{c \times dt}{x}
\end{equation}

\begin{figure}
\centering
\includegraphics[width=.9\textwidth]{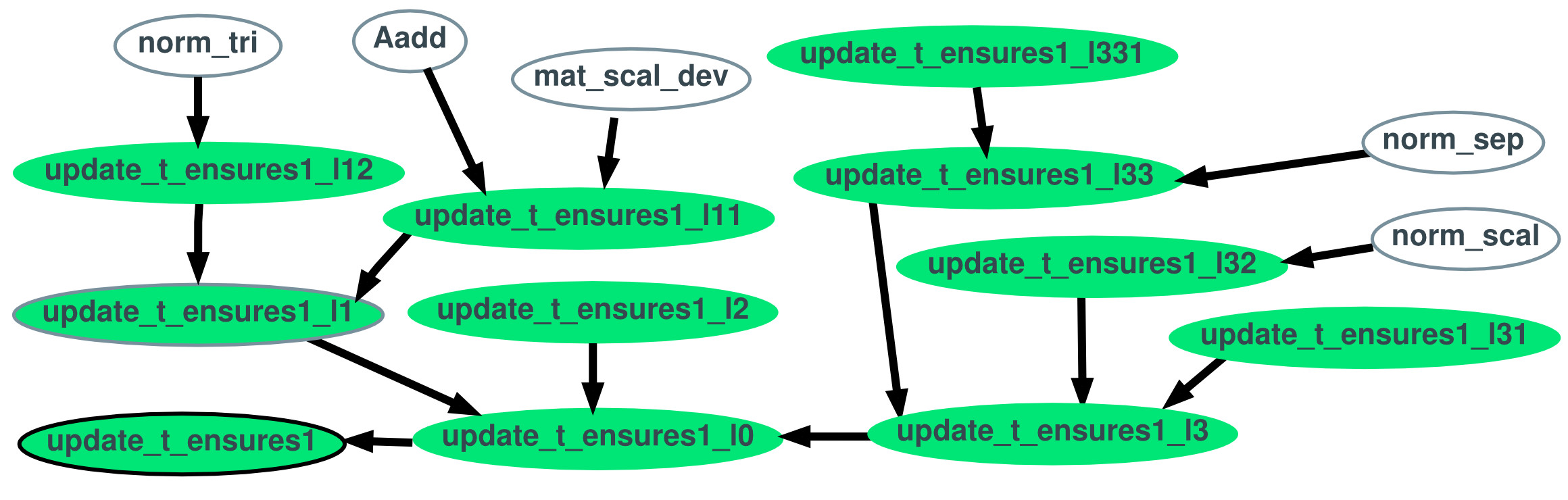}
\caption{Proof tree for~\eqref{eq:update_t_ensures_1} (In green, proven goal, in white, axioms)}
\label{fig:tree}
\end{figure}

Each lemma depends on other lemmas, and so do contracts,
all this logic results shapes a proof tree. The one for the first
ensures of \code{update\_t} is presented in Fig.~\ref{fig:tree}.

These small steps are usually bigger than the application of a tactic in
a proof assistant. SMT solvers are also more resilient to a change
in the source code. For these two reasons we decided not to write proofs
directly within a proof assistant but to use SMT solvers. Moreover,
all these intermediate
lemmas are eventually automatically  generated by the autocoding framework.

%% file: 6-results.tex
\section{Experimentations}
\label{sec:results}

We implemented the presented approach: considering a provided LP problem, a
source file is generated as well as a considerable amount of function contracts
and additional lemmas -- all expressed in ACSL~\cite{acsl}, the annotation
language of the \textsf{Frama-C} verification platform. The function contracts
specify the functional requirements of this custom-code and instance-specific
solver: it computes a feasible and optimal solution in a predictable -- and
known -- number of iterations.

Now, this specific code has to be validated before being used on the
embedded device. We rely on Frama-C to perform the formal verification
process. Each generated annotation becomes a proof objective and is
submitted to an SMT solver, \textsf{Alt-Ergo}~\cite{altergo}, for
verification. Thanks to our introduction of all these intermediate
lemmas, each local lemma is proved as well as the global function
contracts.

Note that the generated code is a direct implementation of the chosen
primal IPM for LP. The main concern here is the proof.  We present
here our experimentations. First, we address issues with respect to
the provers (Frama-C and Alt-Ergo) and how we tuned the code and proof
generation to ease the proofs. Second, we present the results in terms
of goals proved or not-yet-proved and the time required to perform the
proofs.

\subsection{Limitation of the approach: Asserted lemmas.}

The verification is performed thanks to two sets of axioms we
developed, supporting (1) the definition of matrices in ACSL or (2)
properties of operators in linear programming.  While the first set
describing matrices commutation or norm positivity could be proved, it
is an additional work outside of the scope of our contribution. The
second set is more challenging to prove and corresponds to the
axiomatization of some theorems from \cite{nesterov94}. Their
definition as axioms was carefully addressed and the number of such
axioms minimized.



We deliberately choose to ignore:
\begin{itemize}
\item the proof of soundness of the Cholesky resolution;
\item the validation of the gradient or Hessian computations;
\item some vector and matrix related properties: eg. matrix multiplication
  associativity or norm positivity (scalar product);
\item more involved optimization related theorems used within the proof of
  optimality, and developed in~\cite{nesterov}.
\end{itemize}

All these axioms are potential flaws for the verification. One
solution could be to translate them within a Theorem prover such as
\textsf{Coq}~\cite{Bertot2004} and proving them instead of assuming
them.  
Translating matrices to a theorem prover would also have the advantage to check
the matrix axiomatic we wrote. This could be done by providing a
bijection between the translation of our matrices and matrices defined
in \textsf{Coq} libraries like \textsf{SSReflect}~\cite{ssreflect}.




Last, the analytic center has to be provided, at least within the
$\beta$-neighborhood, which is performed automatically through a
dedicated algorithm in our experiments. While its existence cannot be
guaranted in theory, its computation in MPC context can be eased
(cf. Remark 2).

\subsection{Results: Scalability of proof process}



While the custom-code itself ought to be executed on an embedded system with
limited computation power, the verification of its soundness can be performed on
a regular desktop computer.

While all contracts and lemmas were proved, except the set of axioms
we assumed (cf. previous Section), the time and therefore the
scalability of the proof process largely depend on the problem size.

We provide here some figures regarding the total time required to perform the
proofs. The verification has been performed on 2.6GHz computer running Linux
with 4GB of RAM. Alt-Ergo does not parallelize computations.

\paragraph{Total computation time.}

The total proof time directly depends on the size of the input
problem. This is expected since a $n \times m$ problem will manipulate
arrays of that size and express properties over that many variables.

We used our framework on a large set of LP problems of various sizes.
The boundedness constraint of the feasible set
(cf. Sec.~\ref{sec:prelim}) has been artificially enforced by adding a
hypercube constraint. Therefore problems over $n$ variables have at
least $2 \times n$ constraints. In practice, we generated about 100
random instances for each pair $(n, m) \in [2, 4]\times [3 \times n, 5
  \times n [$ and perform the proofs with both a 1 second and 10
    seconds timeouts, for each goal.

An interesting outcome is the relative independence of the proof time with
respect to the actual numerical values of the constraints. Since we recorded the
computation time for each instance of a $(n, m)$ pair, we compare the
experimental results.  As an example, for problems of size $(3,9)$, the total
computation time ranges in $[124s, 127s]$. Note that sparse
constraints will generate more optimized code, minimizing the number of
computations, but are likely to generate harder problem for the solvers.

The number of generated goals is impressive: for a problem of size $(n, m)$, we generate automatically $129 + 18 \times n + 3 \times m + 9 \times n^2$ goals.
We can see that both the number of proofs is linear in $m$ but
quadratic in $n$.  Let us now have a closer look at the proof costs
depending on the category of proof.

\paragraph{Scalability of proofs.}

\begin{figure}
  \begin{center}
    \includegraphics[width=\textwidth]{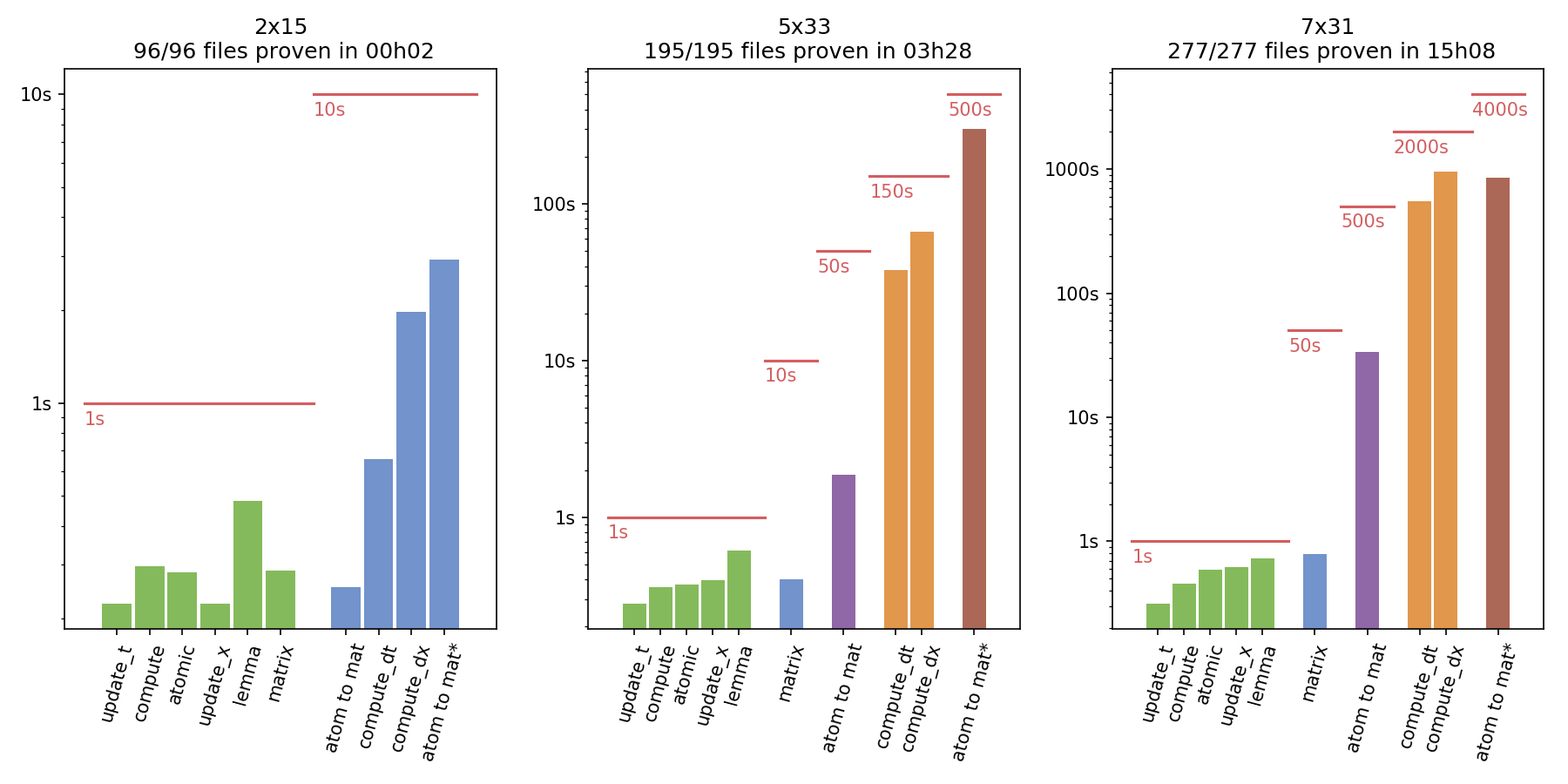}
  \end{center}
  \vspace{-1em}
    Each cluster is described by its average time and the timeout(red lines)
    setting of the SMT solver i.e. an upper bound on the maximum time for each
    goal. As an example, the analyzed instance of size $5\times 33$ produces 195
    goals, all proved. It requires less than one second for the goals of
    categories \textit{lemma}, \textit{atomic}, \textit{update\_t},
    \textit{update\_x} and \textit{compute}; less than 10 seconds for the
    category \textit{matrix} and 50s for the category \textit{atom to mat}. Both
    \textit{compute\_dt} and \textit{compute\_dx} are proved with a timeout of
    150s. The isolated file \textit{atom to mat*} requires a higher timeout of
    500s to achieve all proofs. Note that the average proof time of goals in
    cluster \textit{compute\_dt} is about 40s so only a subset (2 out of 6) of
    the goals require 150s to be proved.

  \caption{Proof-Time required per proof or file category.}
  \label{fig:exp_details}
\end{figure}

In order to precisely keep track of the proof time of each function,
we stored them in separate files and recorded the associated proof
time. Figure~\ref{fig:exp_details} presents such results. Files and
their proof time have been packaged by clusters depending on their
ability to scale, as identified by timeouts. For example one can
identify the following clusters: \textit{lemma} denoting the set of
ACSL lemmas, \textit{atomic} denoting local atomic manipulations of
matrix elements, \textit{matrix} denoting matrix level operation and
\textit{atom to mat} denoting an intermediary results between the last
two. We isolate one of the file from the last cluster : \textit{atom
  to mat*} since it was longer to prove than the others.  Other
categories correspond to individual files implementing higher level
functions of the IPM: \textit{update\_dt}, \textit{update\_dx},
\textit{compute\_dt}, \textit{compute\_dx} and \textit{compute}.

As expected, some properties are difficult to prove automatically. These
scalability issues mainly occur in \textit{compute\_dt}, \textit{compute\_dx}
and the \textit{set\_} and \textit{set\_in} clusters. However, the associated
proofs are not mathematically challenging and the main limitation seems to be
caused by the Frama-C encoding of goals as SMT proof objectives. The future
versions of Frama-C may have better memory model, identifying unmodified
variables in predicates, leading to better proof times of our generated
instances.

 An annotated code, automatically generated by our framework, is accessible at 
 \url{https://github.com/davyg/proved_primal}, annotations can be
 found in the header \url{build_primalTest/code/primalTest.h} which
 itself includes \url{primalTest_contracts.h} the file containing
 all the function contracts.


%% file: 7-related.tex
\section{Related work}
\label{sec:related}

Related works include first activities related to the validation of numerical
intensive control algorithms.  This article is an extension of
\textsc{Wang}\textit{ et al}~\cite{WangJPGFH14} which was presenting annotations for a convex optimization
algorithm, namely IMP, but the process was both manual and theoretical: the code
annotated within Matlab and without machine checked proofs. An other
work from the same authors~\cite{nfm12_2} presented a similar method than ours
but limited to simple control algorithms, linear controllers. The required
theories in ACSL were both different and less general than the ones we are
proposing here.


Concerning soundness of convex optimization, \textsc{Cimini} and
\textsc{Bemporad}~\cite{7907269} presents a termination proof for a quadratic
program but without any concerns for the proof of the implementation itself. A
similar criticism applies to \textsc{T{\o}ndel, Johansen} and
\textsc{Bemporad}~\cite{TondelJB03} where another possible approach to online
linear programming is proposed, moreover it is unclear how this could scale and
how to extend it to other convex
programs. \textsc{Roux}\textit{ et al}~\cite{Roux16,Martin-DorelR17} also presented a way to
certify convex optimization algorithm, namely SDP and its sum-of-Squares (SOS)
extension, but the certification is done a posteriori which is incompatible with
online optimization.

A last set of works, e.g. the work of \textsc{Boldo}\textit{ et al}~\cite{Boldo},
concerns the formal proof of complex numerical algorithms, relying only on
theorem provers. Although code can be extracted from the
proof, the code is usually not directly suitable for embedded system: it is too slow
and requires different compilation steps which should also be proven to have the
same guarantee than our proposed method.


%% file: 8-conclusion.tex
\section{Conclusion and future works}
\label{sec:concl}

This article focuses on the proof a Interior Point Method (IPM) convex
optimization algorithm as used in state-of-the-art linear Model
Predictive Control (MPC). The current setting is the simplest one: a
primal IPM for Linear Programming problems.

In these MPC approaches convex optimzation engines are autocoded from
an instance specification and embed on the cyber-physical system. Our
approach relies on the autocoding fremwork to generate, along the code,
the specification and the required proof artifacrs. Once both the code
and these elements are generated, the proof is achieved automatically
by Frama-C, using deductive methods (weakest precondition) and SMT
solvers (Alt-Ergo).

This is the first proof, at code level, of an IPM algorithm.  Still
the proposed approach is a proof-of-concept and has some identified
limitations. First, we worked with real variables to concentrate on
runtime errors, termination and functional specification and left
floating points errors for a future work. Then, some mathematical
lemmas were asserted since our goal was to focus on the algorithm
itself rather than proving that a norm ought to be positive. The set
of asserted lemmas is still limited and reasonable. Last the setting
is simple: a primal algorithm for LP. The proposed approach is a first
step and can naturally be extended to more sophisticated Interior
Point Methods (IPM): considering primal-dual algorithms, or quadratic
constraints. The limitation is to restrict to IPM algorithms that
guaranty the computation of a feasible solution. For example this does
not include homogenized versions~\cite[\S 11,
  Bibliography]{boyd2004convex} of IPM that do not compute iterates
within the feasible set.

Another outcome is the general approach to deal with the proof of
complex numerical algorithms which are autocoded from an instance
description. Code generation is now widespread for embedded devices,
especially in control, and could be fitted with formal specification
and proof artifacts, to automatize the proof at code level.  This
would require further developments to axiomatize the associated
mathematical theories.



%% file: 00-main.bbl
\begin{thebibliography}{10}

\bibitem{lars}
Blackmore, L.:
\newblock Autonomous precision landing of space rockets.
\newblock National Academy of Engineering, Winter Bridge on Frontiers of
  Engineering \textbf{4}(46) (December 2016)

\bibitem{cvxboyd2014}
Grant, M., Boyd, S.:
\newblock {CVX}: Matlab software for disciplined convex programming, version
  2.1.
\newblock \url{http://cvxr.com/cvx} (March 2014)

\bibitem{hoare69}
Hoare, C.A.R.:
\newblock An axiomatic basis for computer programming.
\newblock Commun. ACM \textbf{12}(10) (1969)  576--580

\bibitem{Floyd1967Flowcharts}
Floyd, R.W.:
\newblock Assigning meanings to programs.
\newblock Proceedings of Symposium on Applied Mathematics \textbf{19} (1967)
  19--32

\bibitem{frama-c}
Cuoq, P., Kirchner, F., Kosmatov, N., Prevosto, V., Signoles, J., Yakobowski,
  B.:
\newblock Frama-c: a software analysis perspective.
\newblock SEFM'12, Springer (2012)  233--247

\bibitem{dijkstra75}
Dijkstra, E.W.:
\newblock Guarded commands, nondeterminacy and formal derivation of programs.
\newblock Commun. ACM \textbf{18}(8) (1975)  453--457

\bibitem{nesterov94}
Nesterov, Y., Nemirovski, A.:
\newblock Interior-point Polynomial Algorithms in Convex Programming. Volume~13
  of Studies in Applied Mathematics.
\newblock Society for Industrial and Applied Mathematics (1994)

\bibitem{acsl}
Baudin, P., Filli{\^a}tre, J.C., March{\'e}, C., Monate, B., Moy, Y., Prevosto,
  V.:
\newblock {ACSL}: {ANSI/ISO} {C} {S}pecification {L}anguage. version 1.11.
\newblock \url{http://frama-c.com/download/acsl.pdf}

\bibitem{nfm12_2}
Herencia{-}Zapana, H., Jobredeaux, R., Owre, S., Garoche, P.L., Feron, E.,
  Perez, G., Ascariz, P.:
\newblock Pvs linear algebra libraries for verification of control software
  algorithms in c/acsl.
\newblock In Goodloe, A., Person, S., eds.: NASA Formal Methods - Forth
  International Symposium, NFM 2012, Norfolk, VA USA, April 3-5, 2012.
  Proceedings. Volume 7226 of Lecture Notes in Computer Science., Springer
  (2012)  147--161

\bibitem{altergo}
Bobot, F., Conchon, S., Contejean, E., Lescuyer, S.:
\newblock Implementing polymorphism in smt solvers.
\newblock In: Proceedings of the Joint Workshops of the 6th International
  Workshop on Satisfiability Modulo Theories and 1st International Workshop on
  Bit-Precise Reasoning. SMT '08/BPR '08, New York, NY, USA, ACM (2008)  1--5

\bibitem{1029817}
Reynolds, J.C.:
\newblock Separation logic: a logic for shared mutable data structures.
\newblock In: Proceedings 17th Annual IEEE Symposium on Logic in Computer
  Science. (2002)  55--74

\bibitem{nesterov}
Nesterov, Y.:
\newblock Introductory lectures on convex optimization : a basic course.
\newblock Applied optimization. Kluwer Academic Publ., Boston, Dordrecht,
  London (2004)

\bibitem{Bertot2004}
Bertot, Y., Cast{\'{e}}ran, P.:
\newblock Interactive Theorem Proving and Program Development.
\newblock Springer Berlin Heidelberg (2004)

\bibitem{ssreflect}
Gonthier, G., Mahboubi, A., Tassi, E.:
\newblock {A Small Scale Reflection Extension for the Coq system}.
\newblock Research Report RR-6455, {Inria Saclay Ile de France} (2016)

\bibitem{WangJPGFH14}
Wang, T., Jobredeaux, R., Pantel, M., Garoche, P.L., Feron, E., Henrion, D.:
\newblock Credible autocoding of convex optimization algorithms.
\newblock Optimization and Engineering \textbf{17}(4) (Dec 2016)  781--812

\bibitem{7907269}
Cimini, G., Bemporad, A.:
\newblock Exact complexity certification of active-set methods for quadratic
  programming.
\newblock IEEE Transactions on Automatic Control \textbf{PP}(99) (2017)  1--1

\bibitem{TondelJB03}
T{\o}ndel, P., Johansen, T.A., Bemporad, A.:
\newblock An algorithm for multi-parametric quadratic programming and explicit
  {MPC} solutions.
\newblock Automatica \textbf{39}(3) (2003)  489--497

\bibitem{Roux16}
Roux, P.:
\newblock Formal proofs of rounding error bounds - with application to an
  automatic positive definiteness check.
\newblock J. Autom. Reasoning \textbf{57}(2) (2016)  135--156

\bibitem{Martin-DorelR17}
Martin{-}Dorel, {\'{E}}., Roux, P.:
\newblock A reflexive tactic for polynomial positivity using numerical solvers
  and floating-point computations.
\newblock In Bertot, Y., Vafeiadis, V., eds.: Proceedings of the 6th {ACM}
  {SIGPLAN} Conference on Certified Programs and Proofs, {CPP} 2017, Paris,
  France, January 16-17, 2017, {ACM} (2017)  90--99

\bibitem{Boldo}
Boldo, S., Faissole, F., Chapoutot, A.:
\newblock Round-off error analysis of explicit one-step numerical integration
  methods.
\newblock In: 2017 IEEE 24th Symposium on Computer Arithmetic (ARITH). (July
  2017)  82--89

\bibitem{boyd2004convex}
Boyd, S., Vandenberghe, L.:
\newblock Convex optimization.
\newblock Cambridge University Press, New York, NY, USA (2004)

\end{thebibliography}
